\documentclass[10pt,conference]{IEEEtran}
\usepackage{tikz}
\usepackage{xcolor}
\usepackage{hyperref}
\usetikzlibrary{quantikz}
\usepackage{subfig}
\usepackage{listings}
\usepackage{booktabs}
\usepackage[utf8]{inputenc}

\usepackage{mathtools}

\usepackage{amssymb}
\usepackage{amsmath}

\usepackage{myStyles/conveniencefunctions}
\usepackage{myStyles/qubitFuns}
\usepackage{myStyles/qutritFuns}
\usepackage{myStyles/blockSpheres}
\usepackage{myStyles/qubitCircuits}
\usepackage{myStyles/qutritCircuits}

\usepackage[backend=biber, style=ieee, url=false]{biblatex}
\begin{document}

\title{Exploring the Potential of Qutrits for Quantum Optimization of Graph Coloring}

\author{
\IEEEauthorblockN{Gabriel Bottrill}
\IEEEauthorblockA{Department of Physics and Astronomy and \\ Department of Computer Science \\ The University of British Columbia \\ Vancouver, Canada \\ bottrill@student.ubc.ca}
\and
\IEEEauthorblockN{Mudit Pandey}
\IEEEauthorblockA{Department of Electrical and \\ Computer Engineering \\
The University of British Columbia\\Vancouver, Canada \\muditp99@student.ubc.ca}
\and
\IEEEauthorblockN{Olivia Di Matteo}
\IEEEauthorblockA{Department of Electrical and \\ Computer Engineering \\
The University of British Columbia\\Vancouver, Canada \\olivia@ece.ubc.ca}
}

\maketitle

\begin{abstract}
Recent hardware demonstrations and advances in circuit compilation have made quantum computing with higher-dimensional systems (qudits) on near-term devices an attractive possibility. Some problems have more natural or optimal encodings using qudits over qubits. We explore this potential by formulating graph 3-coloring, a well-known and difficult problem with practical applications, using qutrits, and solve it using the quantum approximate optimization algorithm (QAOA). Qutrit-based cost and mixer Hamiltonians are constructed along with appropriate quantum circuits using qutrit gates. We run noiseless simulations using PennyLane to compare the formulation against qubit-based QAOA, and analyze the solution quality and resources required. Preliminary results show that the qutrit encoding finds more accurate solutions with a comparable set of hyperparameters, uses half as many qudits, and has a notably smaller circuit depth per layer than an efficient qubit encoding. This work suggests that qutrits may be useful in solving some problems on near-term devices, however further work is required to assess their potential in a noisy environment. 
\end{abstract}


%
\IEEEpeerreviewmaketitle

\section{Introduction}
%
%
%
%
\IEEEPARstart{S}ignificant progress in the design of quantum hardware has taken place over the past decade. However, today's noisy, intermediate-scale quantum (NISQ) devices have many important limitations, including low coherence times, high error rates, a relatively small number of available qubits, and restricted qubit connectivity on some platforms. 

To work within these limitations, a wide variety of Variational Quantum Algorithms (VQAs) have emerged. VQAs leverage trainable (parametrized) quantum circuits in a tightly-coupled optimization loop with classical control hardware. Flagship algorithms such as the variational quantum eigensolver (VQE) \cite{Peruzzo_2014} and quantum approximate optimization algorithm \cite{farhi2014quantum} have been applied in a variety of use cases and demonstrated using multiple hardware modalities. Many open questions remain, for example, regarding their trainability in the presence of barren plateaus  \cite{bar_plat}, parametrized circuit (ansatz) design, and most critically, whether such algorithms will yield a demonstrable quantum advantage on NISQ devices in the near term.

Both VQE and QAOA require domain-specific problems to be mapped (encoded) onto quantum systems, and the design of suitable parametrized circuits and cost functions. The typical encoding to qubits, however, is not always the most efficient or most natural. Many problems would benefit from an encoding into higher-dimensional systems, i.e., \emph{qudits}.

\begin{figure}[htbp]
\centerline{\includegraphics[scale=0.3]{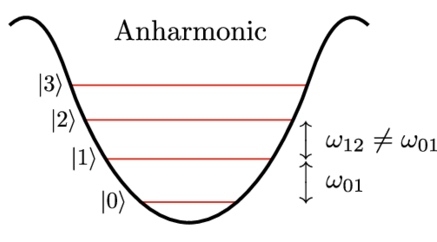}}
\caption{An anharmonic oscillator depicting the energy levels of a quantum system (figure adapted from \cite{anharmonic_oscillator}). Typically, higher energy levels are suppressed to obtain a qubit, however it is possible to use and measure these higher energy states, resulting in a qudit.}
\label{fig:anharmOscillator}
\end{figure}

Some physical implementations, such as superconducting quantum computing platforms, inherently have more levels as shown by \autoref{fig:anharmOscillator}. Hardware providers typically seek to suppress these levels in order to reduce noise and leakage that would negatively affect qubit-based computation  \cite{Smith_2020}. However, some providers are making qudits available at the pulse level, and both hardware and algorithm designers are exploring their use \cite{pulseControl, galda2021implementing}.

For example, work has shown that leveraging higher levels can reduce resources across the compilation pipeline, such as qudit-assisted Toffoli decompositions \cite{gokhale2019asymptotic, litteken22_commun_trade_offs_inter_qudit_circuit}. There have also been recent experimental demonstrations using qutrits to execute actual quantum algorithms \cite{roy22_realiz_two_qutrit_quant_algor}. Other studies have shown that qudit's main source of error comes from decay during measurement \cite{ququad}, for which there are straightforward post-processing models for mitigating this type of error.
Importantly, gate infidelities of qudit implementations are not disproportionately worse than qubit gate infidelities \cite{ququad, roy22_realiz_two_qutrit_quant_algor}. Given these benefits and that the greater available computational space allows for larger problems to be encoded onto current devices, qudits may thus be a crucial piece to finding near-term quantum advantage. 

In this work, we compare qubit and qutrit encodings for solving the graph 3-coloring problem using QAOA. Graph 3-coloring is a common scheduling problem with real-world applications \cite{graph_col}. It is of particular interest in our context as qutrits act as a natural encoding while 3-coloring requires any qubit encoding to work around extra, ``illegal'' states. We implement both methods using PennyLane \cite{pennylane}, a Python library for differentiable quantum programming. As a precursor to this work, we have contributed qutrit simulation capabilities to PennyLane in order to fully support  methods for finding the gradient of qutrit operations using parameter-shift rules. Our work aims to analyze the costs associated with both methods and to explore whether qutrits may provide tangible benefits over qubits for VQAs on NISQ devices.
 
\section{Background}
\subsection{Qutrits}

A $d$-level quantum system can be expressed as a linear combination of $d$ basis states, $
\ket{\psi}=\sum_{k=0}^{d-1} \alpha_k\ket{k}$. In this work we focus on qubits ($d=2$), and qutrits ($d=3$), whose states we write as
\begin{equation}
 \ket{\psi}=\alpha_0\ket{0}+\alpha_1 \ket{1}+\alpha_2 \ket{2}.
\end{equation}
To execute a quantum computation using qutrits, we require a universal set of unitary operations, and a means of performing measurement. The literature contains a number of descriptions for qutrit operations. We choose a natural representation based on rotations using two-dimensional subspaces \cite{di2012elementary}, and Gell-Mann observables, which naturally generalize the qubit Pauli operations and describe a measurement in the qutrit computational basis.

In particular, we use the formalism presented in \cite{di2012elementary}, which defines three rotation matrices analogous to the qubit rotations $RX(\theta)$, $RY(\theta)$, and $RZ(\theta)$. We denote them as $TRX^{(ij)}(\theta)$, $TRY^{(ij)}(\theta)$ and $TRZ^{(ij)}(\theta)$. These operations act as a qubit-like rotation by $\theta$ in the subspace of $\{\ket{i}, \ket{j}\}$ while leaving the third state untouched. As an explicit example, for the $(ij) = (01)$ case, we define
\begin{eqnarray}
TRX^{(01)}(\theta) &=& \begin{pmatrix}
 \cos(\theta/2) & -i \sin(\theta/2) & 0 \\
 -i \sin(\theta/2) & \cos(\theta/2) & 0 \\
 0 & 0 & 1
\end{pmatrix}, \nonumber \\
TRY^{(01)}(\theta) &=& \begin{pmatrix}
 \cos(\theta/2) & - \sin(\theta/2) & 0 \\
  \sin(\theta/2) & \cos(\theta/2) & 0 \\
 0 & 0 & 1
\end{pmatrix},  \label{eq:trs} \\
TRZ^{(01)}(\theta) &=& \begin{pmatrix}
 e^{-i \theta/2} &  0 & 0 \\
  0 & e^{i \theta/2} & 0 \\
 0 & 0 & 1
\end{pmatrix}. \nonumber
\end{eqnarray}

\noindent Operations in the $(12)$ and $(02)$ subspaces are defined similarly. 
An arbitrary single-qutrit operation can be implemented with a sequence of eight such rotation gates, comprising four distinct rotations working in two different subspaces\cite{di2012elementary}\footnote{For example, \cite{di2012elementary} uses the set $\{TRY^{(01)}, TRY^{(02)}, TRZ^{(01)}, TRZ^{(02)}\}$, but as is the case with qubit operations, this choice is not unique.}
The choice of subspaces leads to generalizations of non-parametrized gates, such as analogs of Pauli $X$ that perform addition modulo $3$ on the basis states. The Hadamard gate, which will be required later for QAOA, can be defined in multiple ways \cite{di2012elementary, yeh2022constructing}. One option is the equivalent of a single-qubit Hadamard acting in a two-state subspace. Another version, which we use and define as $TH$, is the qutrit Clifford gate \cite{yeh2022constructing}:
\begin{equation}
    TH = \frac{-i}{\sqrt{3}} \begin{pmatrix}
        1 & 1 & 1 \\
        1 & \omega & \omega^2 \\
        1 & \omega^2 & \omega
    \end{pmatrix} , \quad \omega = e^{2\pi i /3}.
    \label{eq:th}
\end{equation}

The set of observables we measure are the Gell-Mann observables \cite{Gell-Mann}. These are a set of 8 Hermitian observables which generalize the qubit Pauli group. Furthermore, they constitute the Hermitian generators of the unitary operations $TRX$, $TRY$, and $TRZ$ defined in \eqref{eq:trs}. The Gell-Mann observables are enumerated in \autoref{tab:gell-mann}.

\begin{table}[h]
\label{tbl:gellmann}
\centering
\setlength{\tabcolsep}{2.15pt}
\begin{tabular}{c|ccc}\toprule
Pauli & Gell-Mann observable \\
\midrule
\verb+X+ & $\lambda^1= \gOne$ & $\lambda^4= \gFour$ & $\lambda^6= \gSix$ \\
\\ 
\verb+Y+ & $\lambda^2= \gTwo$ & $\lambda^5= \gFive$ & $\lambda^7= \gSeven$ \\
\\
\verb+Z+ & $\lambda^3= \gThree$ & $\lambda^8= \gEight$\\
\bottomrule
\end{tabular}
\caption{Qutrit Gell-Mann observables, grouped by their analogous qubit Pauli operators.}
\label{tab:gell-mann}
\end{table}


For a two-qutrit operation, we choose the ternary addition operator, $TAdd$, which performs  controlled addition modulo 3. The action of $TAdd$ and its adjoint (which we denote as $TSub$) on the qutrit basis states is
\begin{eqnarray}
    TAdd \ket{j} \ket{k} &=& \ket{j} \ket{k + j \hbox{ mod }3},  \\ \label{eq:tadd}
    TAdd^\dagger\ket{j} \ket{k}  &=&  \ket{j} \ket{k - j \hbox{ mod }3} = TSub\ket{j} \ket{k}. \nonumber
\end{eqnarray}

Finally, as we will be using such operations in a variational algorithm, we require that parametrized qutrit gates are differentiable. We can construct parameter-shift rules for $TRX, TRY$ and $TRZ$ based on the differences of eigenvalues of their Hermitian generators:
\begin{eqnarray}
    \frac{\partial U(\theta)}{\partial \theta} &=& \frac{2 + \sqrt{2}}{8} \left(U(\theta + \pi/2) + U(\theta - \pi/2) \right) \label{eq:param-shift} \\ 
    &\enskip& \nonumber  - \frac{2 - \sqrt{2}}{8} \left( U(\theta + 3\pi/2) - U(\theta - 3\pi/2) \right). 
\end{eqnarray}

\subsection{Graph Coloring}
The 3-coloring problem takes as input an undirected graph $G = (V, E)$ and asks if there is a possible coloring of the vertices, $V$, such that no edge in $E$ connects two nodes with the same color.
Although the premise is simple and solutions are easy to check, this problem is NP-complete \cite{graph_col}.
NP-complete problems such as 3-coloring are of particular interest as they all have polynomial transformations into each other while spanning a vast array of important applications.
Finding an algorithm to solve 3-coloring with low error rate would be significant, and as such there is a breadth of work using VQAs to solve this problem \cite{rqaoa, Tabi_2020, photonicQAOA}.

\begin{figure}
    \centering
    \includegraphics[width=\columnwidth]{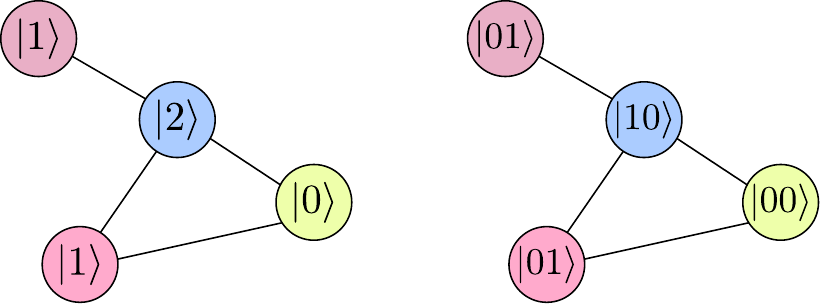}
    \caption{An example encoding of a graph 3-coloring using qutrits (left) vs. qubits (right). The qutrit-based encoding is seemingly more natural given that only a single qutrit is required, and all three basis states are leveraged. The aim of this work is to explore what practical advantages this may offer over the qubit-based version when solving the problem using QAOA, and analyze the tradeoffs involved.}
    \label{fig:graph-3-colors}
\end{figure}

While the solution of graph coloring in general has been explored using qubit-based quantum algorithms \cite{Tabi_2020}, comparatively less work has been done on their solution using qudit-based algorithms. In particular, 3-coloring maps very naturally to a qutrit-based solution, as each color can be mapped to a qutrit computational basis state, as shown in \autoref{fig:graph-3-colors}.

Using qubits for the same problem requires a more complex encoding. Some work has taken a one-hot encoding approach \cite{one_hot}, while others choose a space-efficient encoding that uses $n$ qubits to represent $2^n$ colors \cite{Tabi_2020}.
A space-efficient encoding for 3-coloring requires two qubits per node meaning that a fourth, extra state is present.
This adds complexity to the optimization and must be penalized, which further motivates the use of qutrits.

\subsection{Quantum Approximate Optimization Algorithm}
QAOA finds approximate solutions to combinatorial optimization problems by minimizing the expectation value of a cost Hamiltonian on a parametrized circuit. The cost Hamiltonian must be formulated such that its ground state solutions correspond directly to the analogous solutions of the optimization problem. This cost Hamiltonian, $H_C$, is used to construct a parametrized cost evolution unitary, \begin{equation}
    U_C(\theta)=e^{-i\theta H_C}.
    \label{costUnitary}
\end{equation}
\noindent A mixer Hamiltonian, $H_M$, that does not commute with the cost Hamiltonian is used to define a second parametrized unitary,

\begin{equation}
    U_M(\varphi)=e^{-i\varphi H_M}.
    \label{mixerUnitary}
\end{equation}

\begin{figure}[htbp]
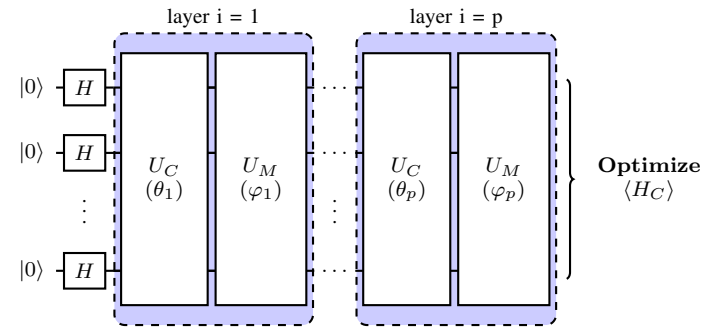

$\QAOA$
\caption{The circuit for QAOA. Each layer contains two trainable parameters, which are optimized classically. The $H$ gate is a standard Hadamard gate in the qubit case, and the $TH$ gate of \autoref{eq:th} in the qutrit case.}
\label{QAOA}
\end{figure}

The algorithm begins by initializing the qudits in the ground state of the mixer Hamiltonian\footnote{This is typically accomplished by applying a Hadamard to every qudit.}. QAOA proceeds by alternating between the two parametrized unitaries of  \autoref{costUnitary} and \autoref{mixerUnitary}, for $p$ layers, as depicted in \autoref{QAOA}.
The parameters of the circuit are then optimized using classical methods. Once trained, the circuit can be run and sampled to obtain solutions with a higher probability.

QAOA has been the subject of interest for solving graph-theoretic problems across multiple domains, such as max-cut. 
Its accuracy increases with the number of layers, however this increases the circuit depth, giving low-depth evolution circuits an obvious advantage \cite{farhi2014quantum}.
This motivates finding encodings with a lower depth per layer; we find that a qutrit encoding satisfies this criteria.  We note that an existing work explores using qutrits specifically for the case of single-layer recursive QAOA, but does not consider the more general $p$-layer version \cite{rqaoa} or present any specific circuit implementations or analysis of the qubit-qutrit tradeoffs involved.

\section{Algorithms and implementation}
To compare qutrit- and qubit-based QAOA algorithms, we must first construct circuits based on the QAOA ansatz. This begins by finding a cost Hamiltonian.
\subsection{Cost Hamiltonians}
\subsubsection{Qubit}
Work has already shown multiple qubit-based encodings for 3-coloring. 
We use a modified version of \cite{Tabi_2020}'s space-efficient encoding Hamiltonian for 4-coloring,
\begin{equation}
H_{CE}=\sum\limits_{[v,w] \in \text{E}}\left[Z_{v_1}Z_{v_2}Z_{w_1}Z_{w_2}
+Z_{v_1}Z_{w_1}+Z_{v_2}Z_{w_2}\right],
\label{bHc4coloring}
\end{equation}
\noindent with an added cost to penalize the fourth color \cite{Tabi_2020},
\begin{equation}
H_{CS}=\sum\limits_{v \in \text{V}}\left[Z_{v_1}Z_{v_2}-Z_{v_1}-Z_{v_2} \right]
\label{bHcNo4}.
\end{equation}

The expectation value of \autoref{bHc4coloring} is minimized for eigenstates representing good colorings, and is higher for eigenstates representing bad colorings. The expectation value of   \autoref{bHcNo4} is low when the qubit pair that makes up each node is in a valid state and high when they are in the $\ket{11}$ state.  The Hamiltonian in \autoref{bHcNo4} will thus be referred to as the suppression term as it suppresses the $\ket{11}$ state.
The entire cost Hamiltonian is 
\begin{equation}
\label{bHc}
\begin{array}{c}
H_C= \sum\limits_{[v,w] \in \text{E}}\left[Z_{v_1}Z_{v_2}Z_{w_1}Z_{w_2}
+Z_{v_1}Z_{w_1}+Z_{v_2}Z_{w_2}\right]\\
+\alpha\sum\limits_{v \in \text{V}}\left[Z_{v_1}Z_{v_2}-Z_{v_1}-Z_{v_2} \right],
\end{array}
\end{equation}
\noindent where $\alpha$ is a real-valued coefficient defining the magnitude of this cost.

\subsubsection{Qutrit}
Often when one is making a cost Hamiltonian for a qubit-based encoding it makes sense to find an equivalent version in terms of binary variables and then translate the solution to the qubit domain.
Unfortunately, this does not make as much sense in the case of qutrits as the qutrit operations and Gell-Mann observables have different behaviour than ternary logic.
However, it is possible to construct an appropriate cost Hamiltonian by using combinations of the Gell-Mann observables $\lambda^3$ and $\lambda^8$, which are analogous to the qubit Pauli $Z$ used in the qubit cost Hamiltonians.
A valid cost Hamiltonian was found through a computational search:
\begin{equation}
H_C=\sum\limits_{[v,w] \in \text{E}}(\lambda^3_v \lambda^3_w)+(\lambda^8_v \lambda^8_w).
\label{qC}
\end{equation}
This cost is promising as not only is it simple, but all six permutationally equivalent colorings have the same expectation value (and similarly for invalid solutions) so solutions are not biased towards any particular assignment of colors.

\subsection{Mixer Hamiltonians}
\label{circDecomp}
A mixer Hamiltonian that does not commute with the cost Hamiltonian must also be chosen. The mixer is a hyperparameter of the algorithm and there are many choices. As an infinite number of mixers are possible, it can be advantageous to create simple low-depth circuits and then convert them back to the Hamiltonian.

\subsubsection{Qubit}
A selection of mixers were tested by running the qubit algorithm using each mixer on a consistent set of graphs with eight nodes of varying connectivity. Probabilities of sampling a solution were plotted and the best candidate, the mixer where an $RX$ gate is applied to each qubit, was selected to continue comparisons against the qutrit algorithm.

\subsubsection{Qutrit}
To construct a mixer for qutrits, it is possible to apply just one $TRX$ subspace rotation, however the resulting Hamiltonian shares most eigenstates with the cost Hamiltonian.
Instead, we use all three subspace rotations sequentially, guaranteeing the minimal amount of shared eigenstates without using entangling gates. While other constructions are possible and could be explored, this mixer has the most natural form (and is analogous to the qubit one).

\subsection{Unitary Circuit Decompositions}
\label{circDecomp}
To execute QAOA, the evolutions under the cost and mixer Hamiltonians must be converted into elementary gates. In the qubit case, the gate set is $RX$, $RZ$ and $CNOT$; for the qutrit case, it is $TRX$ and $TRZ$ for various subspaces, and $TAdd$.

\subsubsection{Qubit}
\begin{figure}[htbp]
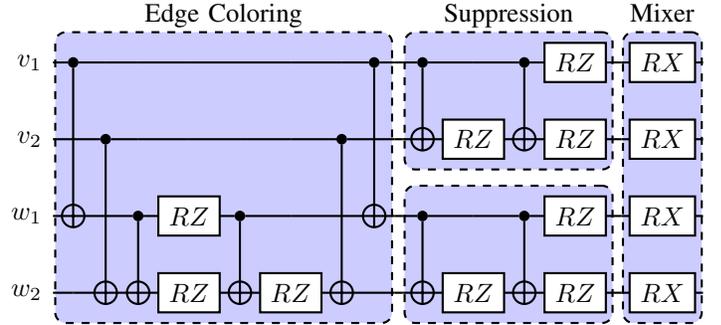

\centerline{\qubitCostFullGateGroups}
\caption{The qubit-based circuit decomposition of one QAOA layer used for a graph with one edge. The Edge Coloring term is applied for each edge in a graph, while the suppression and mixer terms are applied only once for each node.}
\label{fig:costQubit}
\end{figure}

To implement evolution under the cost Hamiltonian we treat the edge coloring Hamiltonian in \autoref{bHc4coloring} and the suppression Hamiltonian of \autoref{bHcNo4} separately creating, 

\begin{equation}
    U_{CE}(\theta)= \exp(-i\theta H_{CE}),
\label{eqn:qubitUnitaryEq}
\end{equation}
and,
\begin{equation}
U_{CS}(\gamma)= \exp(-i\gamma \alpha H_{CS}),
\label{eqn:qubitUnitaryEq}
\end{equation}
respectively. These unitaries are defined in \autoref{eqn:qubitUnitaryEq} and must be applied like this as the coloring Hamiltonian applies to edges, while the suppression Hamiltonian applies only once for each node. $U_{CS}$ is easily decomposed and shown in \autoref{fig:costQubit}. We make use of circuit identities for the Pauli gadgets that implement the typical cascades of $CNOT$s in logarithmic depth (this has the further positive effect of cancellation of some $CNOT$ gates in subsequent terms and the suppression term). The resultant Edge coloring term $U_{CE}$ is shown in \autoref{fig:costQubit}.

Applying both the cost and mixer unitaries will result in a circuit with depth approximated by $6m+5$, where $m$ is the maximum node degree in the graph. This is assuming a fully-connected device layout, and a best case scenario graph allowing for maximum parallelization. 
The circuit corresponding to a graph made of two nodes connected by an edge is shown in \autoref{fig:costQubit}. 

\begin{figure*}[htbp]
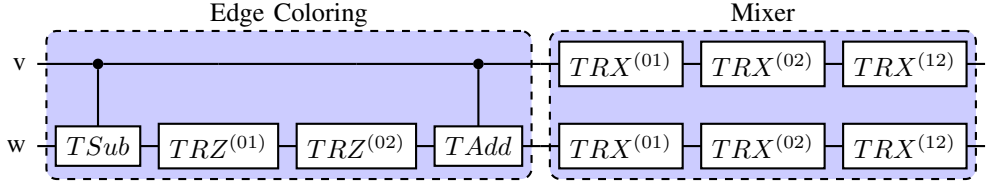

\centerline{\qutritUnitary}
\caption{The qutrit-based circuit decomposition of one QAOA layer for a graph with one edge. The Edge Coloring term is applied for each edge in a larger graph and the mixer term is applied only once for each node.}
\label{fig:costQutrit}
\end{figure*}

\subsubsection{Qutrit}
We hypothesized that the structure of the unitary over one edge would be similar to the unitary $\exp(-i\theta Z_1Z_2)$. 
Thus we explored circuit structures consisting of two entangling gates surrounding rotation gates applied to the target wire.
$TAdd$, unlike $CNOT$, is not self adjoint so 
the two entangling gates are $TAdd$ and $TSub$.
The rotation gates are discovered by considering the diagonal matrix
\begin{equation}
    U_{C\text{ qutrit\_edge}} = \exp(-i\theta (\lambda^3_1\lambda^3_2 + \lambda^8_1\lambda^8_2))
\label{qutritUnitaryEq}
\end{equation}

\noindent whose entries are $e^{-i\theta\frac{4}{3}}$ and $e^{i\theta\frac{2}{3}}$. Choosing rotations
$TRZ^{(01)}(\frac{4}{3}\theta)$ and $TRZ^{(02)}(\frac{4}{3}\theta)$ results in the correct structure. 
The full circuit, including entangling gates, is shown in the Edge Coloring of \autoref{fig:costQutrit}.

Like the qubit circuit, it is worth considering the depth of the circuit, which will be proportional to $m$, the highest degree of any node in the graph (again assuming a fully-connected device and a best case selection of graph).
\footnote{Finding a maximally parallelized circuit for a graph is an NP-hard problem as minimal $k$-coloring can be reduced to it in polynomial time.} 
Depth of a well-parallelized circuit will be close to $4m + 3$ however, the connectivity of other nodes will affect this.

\subsection{Graph Construction}
\label{sec:graphConstruction}
The algorithms were tested on small non-isomorphic 3-colorable graphs of varying connectivity. The graphs were generated using Nauty \cite{MCKAY201494} and a recursive algorithm was used to select only 3-colorable graphs.
Also, larger 3-colorable graphs were generated to find entangling gate counts using a method similar to the algorithm described in \cite{rqaoa}. The algorithm works by splitting the nodes into three sets and randomly connecting nodes together, creating a tripartite graph.



\subsection{Optimization Parameters}

In order to train QAOA, there are a significant number of hyperparameters that must be chosen. This includes the optimization algorithm, number of optimization steps, step size, number of QAOA layers ($p$), and the value of the suppression cost $\alpha$.
The Adam optimizer was chosen as it is an ensemble method that outperforms many other optimizers.
Currently, the simulation of qutrit circuits in PennyLane is not as efficient as for qubit circuits; as a result, the number of optimization steps for qutrits was capped to 50. This resulted in convergence of the cost expectation value to within 0.01 in small graph instances.

The qubit-based algorithm could not converge in 50 steps, so we removed the step limit and allowed for convergence to 0.001 or a minimum of 200 steps.
This allowed for more reasonable solutions as with just 50 steps true colorings were not sampled for qubits, which biases the analysis against the qubit algorithm.
Also, the suppression cost $\alpha=2$ was decided as it is proportional to the coloring cost and gave better results than larger values.
Finally, testing showed that separating the cost unitary trainable parameter into one parameter for the edge coloring unitary and one parameter for the suppression unitary generated significantly better results. 
Therefore this modified version of QAOA was used.

\section{Results}
Our algorithms were implemented using the development version of PennyLane 0.32.0-dev \cite{pennylane}, with the JAX backend \cite{jax2018github} to speed up gradient calculation. Simulations were run on a desktop computer with 64 GB of RAM and an i7-13700KF processor with a RTX 4090 graphics card. The code and numerical data is available at \cite{Bottrill_QTCorGI}.

\begin{figure}[htbp]
\centerline{\includegraphics[scale=0.5]{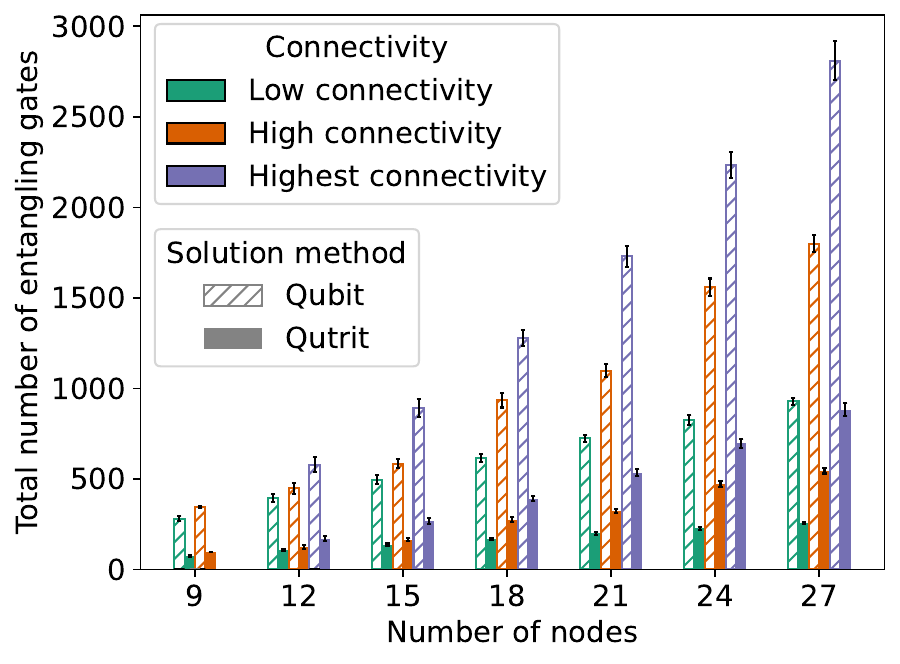}}
\caption{Comparison of the number of entangling gates for QAOA with $p=3$. Low connectivity refers to most nodes having degree 3, whereas high connectivity means most nodes have degree $n/3$, and highest connectivity means most nodes have degree $2n/3$.}
\label{fig:entGateCount}
\end{figure}

To assess the quantum resource requirements of the algorithm, sets of up to 20 non-isomorphic 3-colorable graphs were compiled. 
Each set is constructed with a set number of nodes ($n=9$ to $n=27$, in steps of 3) and connectivity using the second method  outlined in \autoref{sec:graphConstruction}. 
Circuits for 3-layer QAOA are constructed, and the average number of entangling nodes for both circuits is shown in \autoref{fig:entGateCount}.
As already seen from \autoref{fig:costQubit} and \autoref{fig:costQutrit}, the qubit circuits require many more entangling gates than their qutrit counterparts.

\begin{figure}[htbp]
\centerline{\includegraphics[scale=0.5]{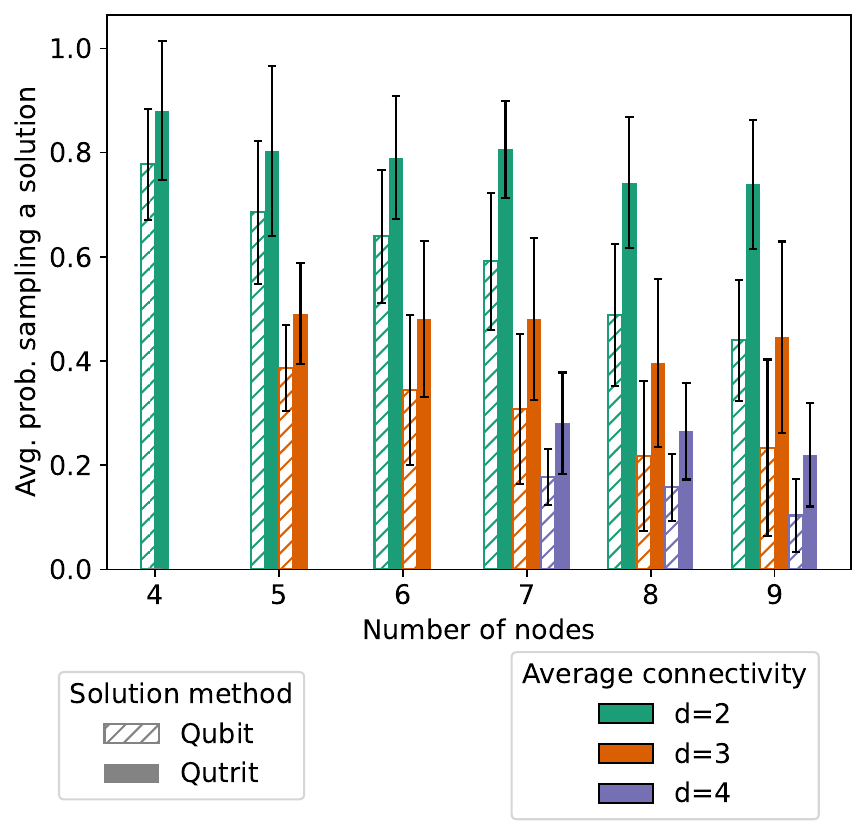}}
\caption{Comparison of the average probability of sampling the correct solution using three QAOA layers of multiple graphs with $n$ nodes. Average connectivity is the average number of edges per node. All samples of the qubit algorithm that include the invalid state are removed. When not post-selected the margins are closer, however this is expected as it would be a four-coloring. The error bars shown correspond to the standard deviation.}
\label{fig:probSample}
\end{figure}

Next, QAOA is simulated to compare the quality between the qubit/qutrit versions. Non-isomorphic sets of up to 20 3-colorable graphs are produced (some smaller graph sizes have less than 20 non-isomorphic graphs), but with fewer nodes ($n=4, \ldots, 9$) due to memory constraints. 
We train the algorithm with $p=3$ QAOA layers on each graph and determine the probability of sampling a correct solution after training.
This probability is averaged for each set of graphs and is displayed in \autoref{fig:probSample}. These results showed that the qutrit-based algorithm led to a higher probability of sampling a correct solution, despite the optimization algorithm not having converged to the same degree as the qubit-based version.

\begin{figure}[htbp]
\centerline{\includegraphics[scale=0.5]{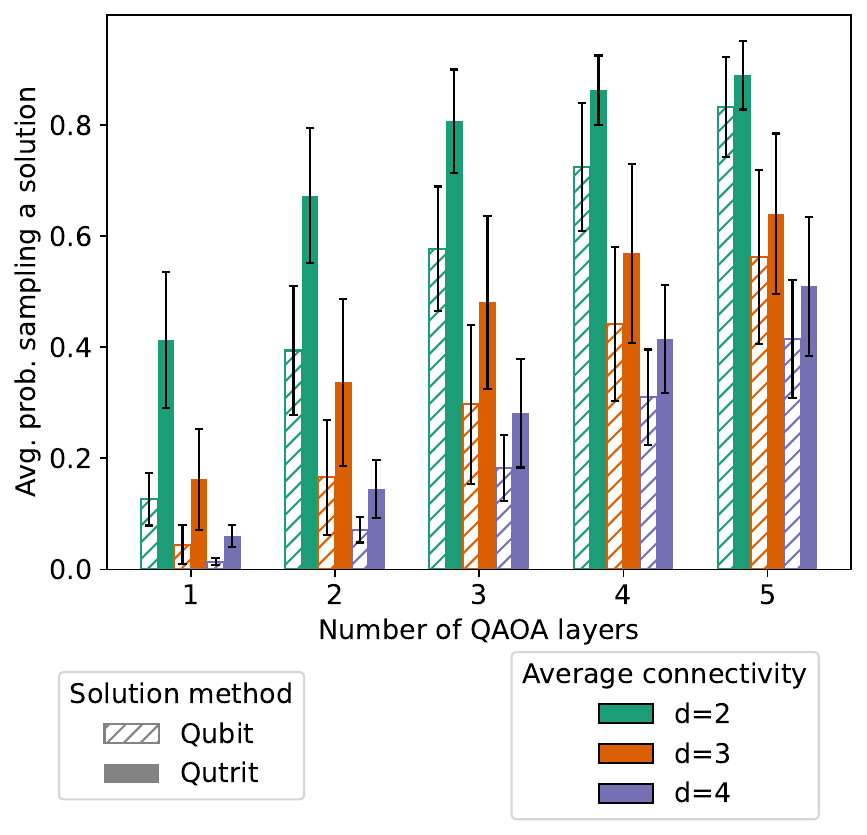}}
\caption{Comparison of the average probability of sampling the correct solution of five graphs for $n=7$ nodes with a varying number of QAOA layers, $p$. All samples of the qubit algorithm that include the invalid state are removed.}
\label{fig:probSampleLayers}
\end{figure}

Finally, we select sets of 20 graphs with $n=7$ nodes and increasing connectivity between sets, and run both algorithms with varying numbers of layers, $p$.
The average probability of sampling a solution is shown in \autoref{fig:probSampleLayers}, where it can be seen that while for both versions of the algorithm the solution quality improves with increasing $p$, the qutrit algorithm consistently outperforms the qubit algorithm. This is notable as it suggests that we can obtain better results using not only fewer qudits, but also fewer layers of circuits requiring fewer resources.

\section{Conclusion and Future Work}

We presented a formulation of QAOA for solving graph 3-coloring using qutrits.
We tested this against an efficient qubit embedding of the problem.
Our results show a significant decrease in cost metrics including entangling gates per layer and circuit depth.
Noiseless simulations show that the qutrit algorithm has a higher probability of sampling a correct solution, and can do so with fewer QAOA layers.
This indicates that the qutrit algorithm may be less susceptible to noise
which is important given that qutrit implementations have higher error rates \cite{ququad, roy22_realiz_two_qutrit_quant_algor}.
These metrics are promising and suggest qutrit-based variational quantum algorithms may be better for solving some combinatorial problems.

More work must be done to perform a thorough analysis of the tradeoffs involved. For instance, further tuning of the optimization hyperparameters is necessary to ensure a consistently fair comparison between the two encodings. Subsequent work will also include simulations with noise, as well as running these algorithms on a real device. Comparing the simulations in the presence of noise will be interesting in particular because, while the qutrit circuits require fewer resources, the parameter-shift rules call for twice as many circuits to be executed to evaluate the gradients for the qutrit case. An analysis of the relative amounts of noise in qubit vs. qutrit gates on hardware is essential. The trainability of the ansatze themselves is also of interest to study, and the algorithm can be tested on more graphs with different properties (for example, graphs that are not perfectly 3-colorable).

Finally, because minimum $k$-coloring is a more common problem, it is worth considering other qudit implementations. For instance, in 3-coloring, the embedding into 2 qubits leads to extra states that must be suppressed; in 8-coloring this would instead occur in the qutrit case. Depending on $k$, a combination of systems with different dimensions may prove valuable to explore, should their realization in hardware become widely available.



%



\section*{Acknowledgments}
GB and MP acknowledge funding from the NSERC CREATE in Quantum Computing Program, grant number 543245. ODM acknowledges funding from NSERC RGPIN-2022-04609, the Canada Research Chairs program, and UBC. All authors wish to thank the PennyLane team at Xanadu for their time spent in discussions and code review during the development of the qutrit functionality.

\newpage

\ifCLASSOPTIONcaptionsoff
  \newpage
\fi



%
\printbibliography

%









\end{document}